**Applying the estimands framework to non-inferiority trials: guidance on choice of hypothetical estimands for non-adherence and comparison of estimation methods**

Katy E Morgan[1], Ian R White[2], Clémence Leyrat[1], Simon Stanworth[3], Brennan C Kahan[2]

[1] Department of Medical Statistics, London School of Hygiene & Tropical Medicine, London, UK

[2] MRC Clinical Trials Unit at UCL, London, UK

[3] Radcliffe Department of Medicine, University of Oxford, Oxford, UK

*Correspondence to: Brennan Kahan, b.kahan@ucl.ac.uk

**Abstract**
A common concern in non-inferiority (NI) trials is that non-adherence due, for example, to poor study conduct can make treatment arms artificially similar. Because intention-to-treat analyses can be anti-conservative in this situation, per-protocol analyses are sometimes recommended. However, such advice does not consider the estimands framework, nor the risk of bias from per-protocol analyses. We therefore sought to update the above guidance using the estimands framework, and compare estimators to improve on the performance of per-protocol analyses. We argue the main threat to validity of NI trials is the occurrence of "trial-specific" intercurrent events (IEs), that is, IEs which occur in a trial setting, but would not occur in practice. To guard against erroneous conclusions of non-inferiority, we suggest an estimand using a hypothetical strategy for trial-specific IEs should be employed, with handling of other non-trial-specific IEs chosen based on clinical considerations. We provide an overview of estimators that could be used to estimate a hypothetical estimand, including inverse probability weighting (IPW), and two instrumental variable approaches (one using an informative Bayesian prior on the effect of standard treatment, and one using a treatment-by-covariate interaction as an instrument). We compare them, using simulation in the setting of all-or-nothing compliance in two active treatment arms, and conclude both IPW and the instrumental variable method using a Bayesian prior are potentially useful approaches, with the choice between them depending on which assumptions are most plausible for a given trial.

# 1. Introduction

Non-inferiority (NI) trials aim to show a new treatment is not worse than a standard treatment by more than a pre-defined amount (the non-inferiority margin) [1-3]. NI trials are often used in settings where a new treatment may not improve outcomes compared to a standard treatment but is expected to have other benefits such as reduced cost or an improved safety profile.

A longstanding concern in NI trials is that non-adherence due to poor trial conduct can make treatment arms appear more similar than they would be in practice [2, 4-7]. This artificial similarity can increase the risk of declaring non-inferiority when using intention-to-treat (ITT) analyses, even when the new intervention is worse than the standard treatment. For these reasons, major guidelines have historically recommended that ITT analyses be supplemented with per-protocol analyses which exclude non-adherent participants, as this analysis is assumed to be less affected by deviations due to poor study conduct [6, 7]. However, per-protocol analyses do not correspond to a well-defined treatment effect and can be biased due to post-baseline exclusions. Importantly, the bias can either increase or decrease the risk of falsely declaring non-inferiority, depending on the pattern of protocol deviations [8-10]. Attention in recent years has therefore focussed on identifying more suitable estimators which rely on less stringent assumptions [5, 6, 11, 12].

However, with the recent publication of the ICH-E9(R1) addendum, there is growing recognition that investigators should start with the estimand (the treatment effect they wish to estimate), and then choose an estimator aligned to this estimand [13-19]. Thus, there is urgent need to update the standard guidance on analyses of NI trials based on the estimands framework, and to identify appropriate estimators for the chosen estimands. A key component of defining an estimand is specifying how intercurrent events (post randomisation events which affect the interpretation or existence of outcome data, such as non-adherence or treatment discontinuation) are handled. An ITT analysis typically corresponds to an estimand where all intercurrent events are handled using a *treatment policy* strategy, where the event is taken to be part of the treatment condition and thus considered irrelevant. However, this strategy may not always reflect the most important clinical question. Further, it is less clear what estimand strategy a per-protocol analysis corresponds to, whether additional estimands would always be necessary in NI trials, or how best to estimate the appropriate estimands.

Given the uncertainty around both the appropriate application of estimands to NI trials and the most appropriate estimators, we sought to (i) discuss how the estimands framework can be applied to non-inferiority trials; and (ii) compare different methods of estimating hypothetical estimands for NI trials with non-compliance in two active treatment arms. The paper is structured as follows: in section 2 we provide a motivating example, and in section 3 we give recommendations for applying the estimands framework to non-inferiority trials. In section 4 we provide a formal definition of the recommended estimand using the potential outcomes framework, and in section 5 we provide an overview of different estimators that could be used for the recommended estimand. In sections 6 and 7 we provide the methods and results of a simulation study evaluating the different estimators, and in section 8 we provide a re-analysis of our motivating example. We conclude in section 9 with a discussion. The primary focus of this paper is on choice of estimands and choice of estimators for non-inferiority trials, thus we do not discuss issues such as how the non-inferiority margin should be chosen (except to note that such choices should involve consideration of the estimand).

## 2. Motivating example: the TOPPS trial

This work was motivated by the TOPPS (Trial of Prophylactic Platelets) trial, which two authors were involved in (BCK, SS) [20]. TOPPS was a randomised non-inferiority trial comparing two different platelet transfusion policies in patients with hematologic cancers. It assessed whether a non-prophylactic transfusion policy (new treatment; patients only received a platelet transfusion if they showed signs of any bleeding) was non-inferior to a prophylactic transfusion strategy (standard treatment; patients received a platelet transfusion if their platelet count dropped below $10 \times 10^9$ per litre) to prevent major bleeding. The primary outcome was occurrence of at least one WHO grade 2-4 bleed within 30 days of randomisation. The non-inferiority margin was a difference of 15 percentage points, meaning that a non-prophylactic strategy could be considered acceptable for use in practice if it did not increase the number of patients experiencing a bleed by more than 15 percentage points. The main perceived benefits of a non-prophylactic approach were lower risk of transfusion related adverse events and substantial cost savings.

The main intercurrent event was deviation from the allocated transfusion policy by administering a platelet transfusion against protocol. The primary analysis followed an ITT strategy, and was supplemented with a secondary per-protocol analysis which excluded participants who had at least one deviation to their allocated transfusion policy.

However, contrary to conventional wisdom, the per-protocol analysis was in fact less conservative than the ITT analysis. While the ITT analysis did not support non-inferiority (adjusted difference in percentage points of 8.4, 90% CI 1.7 to 15.2), the per-protocol analysis did show non-inferiority of the non-prophylactic approach (adjusted difference 4.5, 90% CI -3.0 to 12.0) (Table S1). This discrepancy between analyses likely occurred due to confounding in the per-protocol analysis, where a much higher proportion of patients who experienced a bleeding event were excluded from the non-prophylactic group compared to the prophylactic group (Table S2). This result highlights the need to identify and adopt estimators which rely on less stringent assumptions than per-protocol analyses.

## 3. Recommendations for applying the estimands framework to non-inferiority trials

In Table 1, we describe the difference in philosophy around the implications of poor adherence (or other intercurrent events) in non-inferiority trials from a statistical vs. estimands perspective. We argue that non-adherence or protocol deviations themselves are not a threat to the validity of non-inferiority trials. Many such intercurrent events occur in routine clinical practice, and are thus simply something that needs to be defined as part of the estimand based on clinical considerations, as in any other trial design.

Rather, the threat to validity comes from "trial-specific" intercurrent events, which we define as intercurrent events that occur in a trial setting but would not occur in routine clinical practice. Such trial-specific intercurrent events may be due to poor study conduct, but may also occur for reasons beyond the investigators' control. For instance, at the start of the COVID-19 pandemic, many trials faced widespread treatment deviations due to lockdowns or lack of availability of study treatments. Though these deviations reflected usual practice at the time of the trial, they would not be expected to occur to such an extent in the future, and thus can be seen to be trial-specific. Likewise, many trials leave treatment decisions up to the individual clinicians who treat participants. Due to uncertainty over the best choice of treatment at the time of trial initiation, clinicians may deviate more from the protocol during the trial than they would afterwards, once the uncertainty has been addressed.

These trial-specific intercurrent events serve to make treatment arms more similar than they would be in a non-trial setting, thus increasing the risk of declaring non-inferiority when the new intervention is in fact worse than standard treatment. Thus, these intercurrent events require careful handling in the estimand definition in order to avoid spurious conclusions of non-inferiority.

However, a complication is that trial-specific intercurrent events may not always be identifiable as such. For instance, in some trials it is possible that there will be more deviations during the trial than would be seen afterwards, once the uncertainty around the optimal treatment is resolved. However, given there will always be some level of non-compliance in practice, it may be impossible to differentiate between deviations which were trial-specific and those which would have also occurred outside the trial setting.

Our recommended approach to defining estimands in non-inferiority trials therefore depends both on whether trial-specific intercurrent events are likely to be an issue, and if so, whether they can be identified. Our recommendations are given in Table 2. Briefly, if trial-specific intercurrent events are not likely to occur, we recommend that a single primary estimand be defined based on clinical considerations, as in any other trial, and non-inferiority be assessed on the basis of this single estimand.

If trial-specific intercurrent events *are* likely to be an issue and *can* be identified, then we recommend a single primary estimand be specified. The strategies to handle non-trial-specific intercurrent events should be chosen based on clinical considerations, as above. However, trial-specific intercurrent events should be handled using a *hypothetical* strategy (where interest lies in what patient outcomes would have been had the trial-specific intercurrent events not occurred), in order to match the treatment effect that would be observed in routine practice, and to avoid spurious conclusions of non-inferiority based on artificial trial-specific intercurrent events.

Finally, if trial-specific IEs *are* likely to be an issue and *cannot* be identified, then we suggest that two estimands be specified. First, a primary estimand should be chosen under the assumption there are no trial-specific intercurrent events (i.e. that all intercurrent events seen in the trial would also have occurred in practice). Strategies to handle each intercurrent event should be based on clinical considerations, as above. A secondary estimand should also be specified, which uses a hypothetical strategy for any intercurrent events which *may* be trial-specific. For instance, in the TOPPS example, a hypothetical strategy would be used to handle any transfusion-related deviations as it is impossible to distinguish which are trial-specific and which are not. Then, if non-inferiority is demonstrated for both estimands, investigators can be sure it is not a spurious conclusion based on trial-specific intercurrent events.

We note that the guidance above provides a general framework for thinking through how to apply estimands to NI trials, though it may not be appropriate for all trials. For instance, power considerations may make the requirement that non-inferiority be demonstrated for both estimands (the primary and a secondary using a hypothetical strategy) prohibitive, and in these situations it may be sufficient to demonstrate NI for the primary estimand, while using results for the secondary hypothetical estimand to ensure the primary results were not unduly affected by trial-specific intercurrent events (i.e. that estimates from the two estimands are not too different). Similarly, if only a handful of trial-specific intercurrent events are anticipated, then their impact on results may be negligible and specifying a single primary estimand may be sufficient.

### 4. Definition of a hypothetical estimand

We now turn our attention to estimation, and we begin by defining the hypothetical estimand, using potential outcomes notation. We define it in terms of a trial with two active treatments (new treatment vs. standard treatment) with "non-compliance" in both arms, where compliance is all-or-nothing, i.e. either participants receive their allocated intervention or they do not, and there is no switching between the active treatments. Here, we consider non-compliance to be a trial-specific intercurrent event; we note that the estimand definition below could easily be extended to include other types of non-trial-specific intercurrent events which are handled using alternative strategies.

First, let $Y$ represent the observed outcome, $Z$ the treatment allocation ($Z = 0$ if the patient is allocated to the standard treatment group, $Z = 1$ if allocated to new treatment), and $C$ the patient's compliance status for their assigned treatment ($C = 1$ if the patient complied with their assigned treatment, and $C = 0$ if they did not). Thus, participants can receive one of three treatments: standard treatment or new treatment (if they are assigned to that treatment *and* they comply), or no treatment (if assigned to either treatment arm but they do not comply).

Then, $Y^{(Z=1)}$ denotes the participant's potential outcome if assigned to the new treatment, and $Y^{(Z=0)}$ their potential outcome if assigned to standard treatment, and $Y^{(Z=1,C=1)}$ and $Y^{(Z=0,C=1)}$ denote their potential outcomes under actual receipt of each treatment.

The hypothetical estimand can then be defined as:

$$E\big(Y^{(Z=1,C=1)}\big) - E\big(Y^{(Z=0,C=1)}\big) \quad (1)$$

i.e. it is the expected difference in potential outcomes between the new vs. standard treatment in the hypothetical setting where all participants would comply with their assigned treatment.

### 5. Overview of estimators for the hypothetical estimand

In this section we describe different estimators which could be used to target the hypothetical estimand. We focus on the setting of two treatment arms with all-or-nothing compliance in each, and a continuous outcome, but each estimator could be extended to handle other types of intercurrent events (e.g. treatment switching), or to handle interventions where compliance is not all or nothing (for instance, in TOPPS where clinicians could comply with the transfusion policies at some time points but not others). We briefly mention the additional assumptions required for each estimator when they are extended to the situation of time-varying treatments (such as in TOPPS). Example Stata code is provided in Table S3 in the supplementary material to implement these estimators when there is 'all-or-nothing' compliance in each treatment arm.

We first define some additional notation. Let $X$ denote an observed binary baseline covariate, and let $U$ denote an unobserved binary baseline covariate (both $X$ and $U$ are used to define certain estimators in this section, and are also used in the simulation study in section 6). For convenience, we assume a single binary variable for both $X$ and $U$, though this could be extended to multiple variables of different types.

Throughout this manuscript we assume that randomisation has been properly implemented, so that the treatment arms are *exchangeable*. We also make two key assumptions for all estimators listed below: (a) *consistency*, that is that $Y = Y^{(Z=z,C=c)}$ if $Z = z$ and $C = c$, for $z = 0,1$ and $c = 0,1$; and (b) *no interference*, that is that $Y^{(Z=z,C=c)}$ is independent of the $Z$ and $C$ values of other participants [21].

*Intention-to-treat*
In the context of a continuous outcome, this estimator would typically involve applying a linear regression model of the outcome $Y$ on treatment $Z$ to the intention-to-treat population, which includes all participants in the trial, regardless of whether they complied or not. This approach estimates a treatment policy estimand, and therefore will only be unbiased for the hypothetical estimand when the two estimands coincide. This could occur, for instance, if (i) there is no non-compliance; or (ii) potential outcomes under compliance are the same as under non-compliance (i.e. when $Y^{(Z=z,C=1)} = Y^{(Z=z,C=0)}$). When neither of these conditions are true, this estimator will be biased for the hypothetical estimand.

*Per-protocol*
A per-protocol analysis is the same as the ITT approach described above, except that participants who did not comply are excluded from the analysis population. Per-protocol analyses can adjust for baseline covariates, such as $X$, as a covariate in a regression model, in case such covariates act as confounders between the outcome and non-compliance.

For 'all-or-nothing' treatments, the assumptions required for unbiasedness are:
- *Conditional exchangeability*, that is that $Y^{(Z=z,C=1)} \perp Z, C \,|X$. This implies that, conditional on $X$, patients who comply with their assigned treatment are *exchangeable* between treatment arms. This is sometimes referred to as the "no unmeasured confounding" assumption [22].
- No treatment effect heterogeneity across levels of $X$.

The latter assumption is required because the per-protocol analysis provides a weighted average of the estimated treatment effects across levels of $X$, however the weighting used does not necessarily correspond to population weights for $X$. Thus, if the treatment effect varies across levels of $X$, then the per-protocol analysis may upweight or down weight treatment effects from certain levels of $X$ more than it should.

It should be noted that for non-collapsible summary measures, such as an odds ratio, adjustment for baseline covariates can change the estimand from a marginal one to a conditional one, which may not be desirable. However, this is not an issue for differences, and so we do not consider this issue further here.

For time-varying treatments or those with partial compliance (where compliance is defined as meeting some threshold of treatment adherence), per-protocol analyses do not require any additional assumptions for unbiasedness, but the plausibility of the "no unmeasured confounding" assumption becomes much less likely because it needs to hold at each time point, and there may be post-randomisation confounding factors which cannot be adjusted for in the analysis.

*Inverse probability weighting*
For 'all-or-nothing' interventions, inverse probability weighting (also termed "inverse probability of censoring weighting") estimates hypothetical treatment effects by excluding participants who did not comply, and re-weighting participants who did comply according to the inverse of their probability of complying [22-26]. Broadly, the idea is to implicitly impute what outcome data for participants who did not comply would have been under hypothetical compliance, by up-weighting outcome data from comparable participants who did comply.

IPW is implemented in two stages. The first stage is used to estimate the weights to be used in the second stage. This is done separately within each treatment arm, and the weights are defined as:

$$W_{Z=0} = \frac{1}{\hat{P}(C=1|Z=0,X)}$$

And:

$$W_{Z=1} = \frac{1}{\hat{P}(C=1|Z=1,X)}$$

where $\hat{P}(C=1|Z=0,X)$ and $\hat{P}(C=1|Z=1,X)$ are estimated using a logistic regression model applied to each arm separately with the participant's compliance status as the outcome, and baseline covariate(s) $X$ as covariates. Then, $\hat{P}(C=1|Z=0,X)$ and $\hat{P}(C=1|Z=1,X)$ are the participant-specific predictions from the logistic models.

In the second stage, the treatment effect is estimated using a weighted regression model, with outcome $Y$, treatment allocation $Z$, and weights $W_{Z=0}$ (if $Z=0$) and $W_{Z=1}$ (if $Z=1$).

For 'all-or-nothing' treatments, the assumptions required for unbiasedness are [22-26]:
- *Conditional exchangeability*, i.e. that all the relevant $X$ variables have been used to estimate the weights, $W_{Z=0}$ and $W_{Z=1}$, so that $Y^{(Z=z,C=c)} \perp Z, C|X$
- The association between $X$ and compliance status $C$ has been correctly specified to estimate the weights during stage 1 (i.e. that there is no residual confounding due to misspecification of the confounder-compliance association)
- There is a non-zero probability of complying in each treatment arm for all combinations of the baseline covariates (this is known as the "positivity" assumption)

For time-varying treatments, the IPW approach described above must be extended to deal with time-varying confounding between post-randomisation variables and compliance status (e.g. if post-randomisation blood measurements make non-compliance more likely) [23-26]. This is done by splitting the follow-up period into distinct time-points, and calculating weights for each distinct time-point (based on the inverse probability of remaining compliant at that time-point, conditional on the participant being compliant up to that point); calculation of these weights would include post-randomisation confounders of compliance status and outcomes at each follow-up time-point. IPW does not require any additional assumptions for unbiasedness in this setting, except that the assumptions listed above now include post-randomisation confounding and positivity (i.e. all baseline *and* post-randomisation confounders have been included and correctly modelled, and there is a non-zero probability of remaining compliant at all follow-up time-points for all combinations of covariates given the compliance history). For time-varying treatments, weights may need to be stabilized [26].

*Instrumental variables*

Instrumental variables (IV) is an analysis technique which uses "instruments" to estimate the effect of adhering to treatment [27-31]. An instrument is a variable that is associated with compliance, but not associated with the outcome *except* through its impact on compliance [27, 28, 30]. A major benefit of IV methods is that they do not require the "no unmeasured confounding" assumption, and thus can provide unbiased estimates even when confounding between compliance status and the outcome occurs. However, they make alternative assumptions which may be more or less plausible depending on context.

We define some additional notation. Let $C_0$ denote actual receipt of treatment 0, so $C_0 = 1$ if $Z = 0$ and $C = 1$, and 0 otherwise, and $C_1$ denote actual receipt of treatment 1, so $C_1 = 1$ if $Z = 1$ and $C = 1$, and 0 otherwise. In randomised trials, randomised arm ($Z$) is typically used as an instrument, though as discussed below, some estimators require additional instruments. There are three essential requirements for a variable to be a valid instrument [27, 28, 30] (and further assumptions for an estimator based on IVs to be unbiased for the hypothetical estimand which are discussed below):

1. The instrument must be associated with treatment actually received (e.g. randomisation to treatment $Z = 1$ is associated with patients actually receiving treatment 1, denoted by $C_1$);
2. The instrument has no effect on the outcome Y except through its effect on treatment received, $C_0$ and $C_1$ (this is commonly referred to as the "exclusion restriction" and means that treatment allocation $Z$ does not causally affect outcome Y in participants for whom $C = 0$);
3. The instrument does not share any common causes with the outcome Y (i.e. the association between $Z$ and $Y$ is unconfounded).

Using randomised arm, $Z$, as an instrument typically fulfils assumptions 1 and 3, though the plausibility of assumption 2 requires context-specific knowledge (e.g. it is plausible for many all-or-nothing treatments, but perhaps less plausible for interventions with partial compliance where $C_0$ and $C_1$ are defined as fully adhering to treatment) [30].

One additional assumption required to estimate the hypothetical estimand is *homogeneity*, i.e. that the treatment effect under hypothetical compliance is the same across all compliance levels. Broadly, this implies the quantity $E(Y^{(Z=1,C=1)} - Y^{(Z=1,C=0)})$ is identical for patients who would comply under either treatment assignment; for those who would comply under assignment to one treatment but not the other; or for those who would not comply under assignment to either treatment.

IV estimation can be best explained using a two-stage approach. Without covariates, the two stages are [30]:
1. Stage 1: a linear regression model is fitted for each treatment arm, with receipt of treatment ($C_0$ or $C_1$) as the outcome and allocation ($Z$) as the covariate (so that participants not assigned to treatment $Z = 0$ are included in the model as $C_0 = 0$, and similarly for participants not assigned to $Z = 1$). A prediction for each participant's treatment received status is then obtained ($\hat{C}_0$ and $\hat{C}_1$)

2. Stage 2: a linear regression model is fitted with $Y$ as the outcome, and compliance predictions $\hat{C}_0$ and $\hat{C}_1$ as covariates. An overall estimate of treatment effect is then obtained by contrasting the estimated parameters for $\hat{C}_1$ and $\hat{C}_0$

A challenge for trials with non-compliance in both treatment arms is that, without covariates, $\hat{C}_1$ and $\hat{C}_0$ are necessarily collinear, leading to an unidentifiable model in stage 2. Thus, IV methods for trials with non-compliance in both treatments require additional assumptions to resolve the collinearity issue. We discuss two of these estimators below.

*IV method 1: IV(interaction)*
Fischer *et al* [32] described an IV approach which avoids collinearity between $\hat{C}_1$ and $\hat{C}_0$ by using randomised arm ($Z$) as the first instrument, and then specifying a second instrument based on the interaction between treatment allocation and a baseline covariate ($ZX$). This approach has also been discussed by others [30, 33]. We refer to this approach as *IV(interaction)*. In the present setting, including covariates in the stage 1 model naturally leads to the use of interactions, because $C_1$ is identically zero in $Z = 0$ but may vary with $X$ in $Z = 1$ (and vice versa).

The stage 1 models for this approach are:

$$C_0 = \begin{cases} \alpha^{C0} + \beta_X^{C0} X + e^{C0} & if\ Z = 0 \\ 0 & if\ Z = 1 \end{cases} \quad (2)$$

$$C_1 = \begin{cases} 0 & if\ Z = 0 \\ \alpha^{C1} + \beta_X^{C1} X + e^{C1} & if\ Z = 1 \end{cases} \quad (3)$$

and the stage 2 model is:

$$Y = \alpha^Y + \beta_{C0}^Y \hat{C}_0 + \beta_{C1}^Y \hat{C}_1 + \beta_X^Y X + e^Y \quad (4)$$

where $e^Y$, $e^{C0}$, and $e^{C1}$ are residual error terms, assumed to be normally distributed with mean zero and (co-)variances specified by a 3x3 matrix. We use superscripts $Y$, $C0$, and $C1$ to denote which model each parameter belongs to.

The treatment effect is then estimated as $\hat{\beta}_{C1}^Y - \hat{\beta}_{C0}^Y$ from model (4).

The key idea behind this estimator can be summarised as follows (for full details, see Fischer *et al* [32]): $Z$ and $ZX$ are used as instruments, and for these to be valid instruments they must predict compliance in both treatment arms (the variables $C_0$ and $C_1$). $Z$ predicts both $C_0$ and $C_1$, provided there is some compliance in both treatment groups (i.e. $E[C|Z] > 0$ for $Z = 0,1$); and $ZX$ predicts $C_1$ if $X$ predicts compliance within treatment arm $Z = 1$. Model (4) is identifiable provided the covariates are not collinear: Fischer *et al* showed that this is true provided the predicted treatment compliances $\hat{C}_0$ and $\hat{C}_1$ are not proportional across levels of the baseline covariate, i.e. there is no $k$ such that $\hat{C}_1 = k\hat{C}_0$ across all levels of $X$. A further requirement for the hypothetical estimand is that the baseline covariate $X$ does not moderate the treatment effect directly (i.e. there is no baseline-by-treatment interaction on outcome) [30].

This estimator could be used for time-varying treatments where compliance is defined as meeting some adherence threshold (e.g. compliant for 80% of study days), however the exclusion restriction assumption above (that $Z$ is not associated with outcome Y in participants for whom $C = 0$) is likely to be violated. For instance, in TOPPS it is likely that a patient who followed the transfusion protocol for 28/30 platelet counts is going to receive some benefit from being allocated to that treatment arm.

*IV method 2: IV(Bayes)*

For many non-inferiority trials, information on the effect of standard vs. no treatment will be available from previous trials which have compared the standard treatment against placebo or previous controls. Bond and White [34] therefore described an IV approach which handles collinearity between $\hat{C}_1$ and $\hat{C}_0$ by using a Bayesian framework to put an informative prior on the parameter $\beta_{C0}^Y$ (the effect of the standard treatment vs. no treatment) in the stage 2 model. Even if previous trial information is not available, it may still be possible to identify plausible priors for $\beta_{Z=0}^Y$, for instance based on clinical knowledge. Other parameters use uninformative priors (though could be made informative if desired). We refer to this approach as *IV(Bayes)*.

The stage 1 models for this approach, without adjustment for covariates, are:

$$C_0 = \alpha^{C0} + \beta_Z^{C0} Z + e^{C0} \quad (5)$$

$$C_1 = \alpha^{C1} + \beta_Z^{C1} Z + e^{C1} \quad (6)$$

and the stage 2 model is:

$$Y = \alpha^Y + \beta_{C0}^Y \hat{C}_0 + \beta_{C1}^Y \hat{C}_1 + e^Y \quad (7)$$

where an informative prior is placed on $\beta_{C0}^Y$ in model (7), and uninformative priors on other parameters in the model. The treatment effect is then estimated as $\hat{\beta}_{C1}^Y - \hat{\beta}_{C0}^Y$.

Although this is a Bayesian approach, we still discuss the assumptions required for this estimator to be unbiased for the hypothetical estimand. For a given prior placed on $\beta_{C0}^Y$, an unbiased treatment effect also requires the mean of the prior to be an unbiased representation of the true effect.

Similarly to the *IV(interaction)* method, this approach could be used for time-varying treatments however violations to the exclusion restriction are more likely, which may introduce bias.

## 6. Simulation study methods

We conducted a simulation study to evaluate the estimators described earlier. The primary aim was to evaluate bias both when the estimators' assumptions were fulfilled, as well as when the assumptions were violated. Secondary aims were to evaluate precision and type I error rate of the estimators.

Simulations were based on a two-arm randomised non-inferiority trial with a continuous outcome. Non-compliance occurred in both treatment arms, and compliance was 'all or nothing', i.e. patients either received their allocated treatment or received nothing.

We performed two simulations studies: the first when there is no treatment effect heterogeneity across compliance levels (a core assumption of the IV methods and the per-protocol analysis), and the second which did include treatment effect heterogeneity across compliance levels (indicating a violation in assumptions for the IV and per-protocol estimators).

Full details of the simulation methods, including exact parameter values for all scenarios, are available in the supplementary material. Stata code used to generate data is available in the supplementary material for two scenarios (one each for simulation studies 1 and 2), and code for the other scenarios was identical except for modifications to the input parameters. Below we summarise the key aspects of the simulation study.

*Simulation study 1 (no treatment effect heterogeneity)*
We generated patient outcomes in two steps. First, we generated whether they complied with their assigned treatment, and then we generated their outcome. Their compliance could depend on treatment allocation $Z$, an observed baseline covariate $X$, and unobserved baseline covariate $U$, and the interactions between treatment allocation and either the observed or unobserved baseline covariate. Their outcome could depend on treatment received, and the observed and unobserved baseline covariates. Inclusion of an interaction between $Z$ and $X$ in the model to generate compliance was used to generate measured confounding between compliance status and outcome, while inclusion of an interaction between $Z$ and $U$ was used to generate unmeasured confounding (see Table 3).

We considered five scenarios (labelled A-E, shown in Table S4) in which we varied the sample size, percentage compliance, true value of the estimand, and association between covariates $X$ and $U$ and outcome. Then, for each of these five scenarios, we also considered eight compliance scenarios (labelled 1, 2a-c, 3a-b, and 4a-b; shown in Table 3 and Table S4). This led to a total of 5x8=40 scenarios. We used a non-inferiority margin of -0.3 in all scenarios.

The aims of scenarios A-E were to assess the impact of smaller vs. larger sample sizes, smaller vs. larger degrees of non-compliance, smaller vs. larger associations between covariates and the outcome, as well as impacts on type I error rate vs. power. The aims of compliance scenarios 1-4 were to evaluate the impact of different types of compliance mechanisms, including when there was no measured or unmeasured confounding between compliance and outcome, when there was measured confounding only, unmeasured confounding only, or both.

*Simulation study 2 (treatment effect heterogeneity)*
We used two treatment effect heterogeneity (TEH) scenarios: one in which the treatment effect varied across values of $X$ (which implies it varies across compliance status, as $X$ is strongly associated with compliance in this scenario), and one in which it varied across values of $U$ (which also implies it varies across compliance status; the key difference between these scenarios is that $X$ is observed while $U$ is not). We label these two scenarios *TEH(X) and TEH(U)*.

For *TEH(X)* and *TEH(U)* we generated data so that there was observed and unobserved confounding between compliance and outcome respectively. For both scenarios we varied two factors: the degree of TEH (moderate vs. large), and the difference in compliance between treatment groups (moderate vs. large).

The aim of simulation study 2 was to evaluate how estimators performed when there was treatment effect heterogeneity as well as confounding between compliance status and outcome (either observed or unobserved).

*Estimators*

We implemented five estimators, as described earlier; (i) intention-to-treat; (ii) per-protocol; (iii) IPW; (iv) *IV(Bayes)*; and (v) *IV(interaction)*. All analyses adjusted for the observed covariate $X$ (except for IPW, which used $X$ to estimate weights). Our primary interest was in evaluating IPW, *IV(Bayes)*, and *IV(interaction)*, however we included intention-to-treat and per-protocol for completeness.

For *IV(Bayes)*, we evaluated four different priors for the effect of standard treatment vs no treatment: (a) a well centred, precise prior (i.e. where the prior's mean matches the true mean in the trial, and the prior has a small variance); (b) a well centred, vague prior (where the prior has a large variance); (c) a miscentred, precise prior (where the prior's mean does not match the true mean in the trial); and (d) a miscentred, vague prior. We used these four priors to evaluate the impact of misspecifying the prior's mean in relation to the true mean in the trial, as well as the impact of more vs less precise priors.

We evaluated estimators based on frequentist properties for bias (our main objective), as well as precision, type I error rate, and bias in estimated standard errors (our secondary objectives).

For each estimator, standard errors were calculated using the default approach in Stata. The estimators that use Stata's `regress` command (intention-to-treat and per-protocol) used ordinary least squares estimates of standard errors. The IPW method was implemented using a weighted version of the `regress` command, and the standard errors were obtained from a sandwich estimator. The IV(Bayes) method uses the standard deviation of the posterior distribution as the standard error of the estimator, and the lower end of the 95% credible interval to determine whether non-inferiority is declared or not. The IV(interaction) estimator uses the default variance estimator given in the Stata manual for ivregress [35].

## 7. Simulation study results

### *Simulation study 1 (no treatment effect heterogeneity)*

Full results for each simulation scenario are available in the supplementary material. Because results between scenarios A-E were broadly similar, we present the results for scenario A below. Our focus is on describing results for IPW, *IV(Bayes)*, and *IV(interaction)*, however intention-to-treat and per-protocol results are available in the figures and in the supplementary material.

**Bias in estimated treatment effects**

Results are shown in figure 1 and the supplementary material. As expected, IPW was unbiased except when there was unmeasured confounding between compliance status and outcome. *IV(Bayes)* was unbiased except when the mean of the prior for the effect of the standard treatment vs. no treatment was misspecified compared to the true mean in the trial. However, when the overall compliance rate was the same in each treatment group, *IV(Bayes)* was unbiased even when the prior was misspecified.

*IV(interaction)* was extremely unstable across all scenarios (see supplementary appendix), and results were severely affected by extreme outliers. After removing replications with extreme values, the method performed better, though was still biased for certain scenarios.

**Precision and type I error rate**

Results are shown in figures 2 and 3, and the supplementary material. As expected, IPW led to inflated type I error rates in the same scenarios for which it was biased (i.e. when there was unmeasured confounding), but maintained type I error rates otherwise. In some simulated datasets, the IPW method dropped some observations due to perfect prediction in the logistic regressions used to generate weights. This generally affected only a small number of datasets in each scenario (between 0 and 0.8% for most scenarios; see tables in supplementary appendix) but was as high as 4.1% when compliance was particularly high.

*IV(Bayes)* controlled the type I error rate at close to the nominal level when the mean of the prior was well specified, but resulted in some type I error rates which were too low when the prior was mispecified. This was because treatment effect estimates were biased away from the null, which made a finding of non-inferiority less likely.

*IV(interaction)* led to type I error rates that were far below the nominal level for all scenarios. This was primarily due to extreme bias in estimated SEs.

In general, IPW was more precise than *IV(Bayes)*, and *IV(interaction)* was the least precise, with losses in precision up to 400% in some cases.

**Precise vs. vague priors for *IV(Bayes)* approaches**

Full results are available in the supplementary appendix. Using a precise vs vague prior had no impact on bias in estimated treatment effects. The precision of the two approaches were very similar, however the estimated SEs from the vague prior were substantially biased upwards (often >20%), which led to type I error rates that were below the nominal level in many cases, and reduced power. Overall, we did not find any benefit in frequentist properties to using a vague prior over a precise prior.

*Simulation study 2 (treatment effect heterogeneity)*

Mean estimated treatment effects are shown in figure 4. IPW was unbiased in scenarios where there was no unmeasured confounding. *IV(Bayes)* with a centred prior had a slight bias in all scenarios, which was more pronounced when there was both a large degree of TEH and large differences in compliance between treatment arms. *IV(Bayes)* with a miscentred prior was extremely biased across all scenarios, as was *IV(interaction)* for most scenarios.

## 8. Re-analysis of TOPPS trial

*Methods*

We re-analysed the TOPPS trial to compare the different estimators in practice. We analysed a secondary outcome, the number of days with bleeding. We chose to analyse this instead of the primary outcome described in section 2 because we wanted to compare the analysis methods on a continuous outcome to match our simulation study. Further, this outcome displays similar results to the primary outcome, where the per-protocol analysis led to a smaller estimate of treatment effect than the intention-to-treat analysis.

A full description of methods is available in the supplementary appendix. Briefly, for intention-to-treat and per-protocol we adjusted for a number of baseline variables as potential confounders. For IPW we calculated weights using a combination of baseline and post-randomisation variables. For *IV(interaction)* we fit four separate models, each using a different baseline covariate as an instrument; the interaction between each covariate and treatment allocation on compliance is shown in the supplementary appendix (Table S6). For the *IV(Bayes)* approach we fit four separate models, each using a different prior for the effect of the active control (the prophylactic strategy). Priors were chosen based on our judgement of what was plausible; this was done specifically for the purpose of this re-analysis, and so they were chosen retrospectively after the trial was already complete. For the four priors, we used combinations of small vs large effects and precise vs vague variances.

*Results*

Results are shown in Table 4. For the number of days with bleeding, the ITT and per-protocol analyses had discrepant results; ITT showed a statistically significant increase (difference non-prophylactic vs. prophylactic 0.6 days, 95% CI 0.2 to 1.0, p=0.004) while per-protocol did not (difference 0.4 days, 95% -0.1 to 0.8, p=0.11). IPW and *IV(Bayes)* also demonstrated significant increases in bleeding days; IPW and *IV(Bayes)* with a small prior both showed results similar to ITT, while *IV(Bayes)* with a large prior showed a larger increase in bleeding days (1.2, 95% CI 0.7 to 1.7). Results for *IV(interaction)* were highly variable depending on which baseline covariate was used as the basis for an instrument; estimates ranged between -1.2 and 3.9. One covariate gave an estimate in the opposite direction as the ITT and per-protocol results, another indicated no effect, and one gave an estimate that was about 6.5 times larger than the ITT effect.

## 9. Discussion

Common advice for non-inferiority trials is that ITT be supplemented by per-protocol analyses as protection against the risk of erroneously declaring NI based on a proliferation of protocol deviations which makes treatment arms more similar than they would be in practice. However, there are two issues with this advice: (i) it is based on statistical considerations alone, and does not consider estimands; and (ii) per-protocol analyses do not inherently protect against protocol deviations – as seen in TOPPS, they can actually increase the risk of erroneously declaring NI due to bias from post-randomisation exclusions. In this article we sought to address the above deficiencies by updating the advice in light of recent focus on estimands, and identifying and comparing methods of estimation which improve on per-protocol.

We argue that non-adherence or protocol deviations themselves are not a threat to the validity of NI trials. Such intercurrent events occur in practice, and are thus simply something that needs to be defined as part of the estimand. Rather, the threat to validity comes from trial-specific intercurrent events (those that occur in a trial setting but would not occur in practice, for instance due to poor study conduct). These intercurrent events can serve to make treatment arms more similar than they would be in practice, thus increasing the risk of declaring NI when the new intervention is in fact worse than control. Further complicating the issue is that intercurrent events that would occur in practice cannot always be distinguished from those that would not. In TOPPS, for example, some degree of non-compliance to the transfusion policies would be expected in practice, albeit to a lesser degree than that seen in the trial, but there is no way to distinguish which category any particular deviation falls under.

We therefore suggest for NI trials where trial-specific intercurrent events may be an issue, they be handled in the estimand using a hypothetical strategy. The hypothetical estimand serves as reassurance that a NI conclusion is real and not due to trial-specific issues. The strategies for other, non-trial-specific, intercurrent events could be based on clinical considerations. We note that our advice does not prohibit the use of the hypothetical strategy for non-trial-specific intercurrent events if clinically warranted. We also note that the underlying factors that introduced trial-specific intercurrent events may also affect other aspects of the trial, and warrants careful consideration by investigators.

Estimation of hypothetical treatment effects can be challenging and requires untestable assumptions. Using simulation and a re-analysis of TOPPS, we evaluated several methods that could be used for NI trials with non-compliance in both treatment arms. We found that IPW and *IV(Bayes)* are good options provided their underlying assumptions are fulfilled. Conversely, *IV(interaction)* did not perform well in the scenarios considered in our simulation study, and so we cannot see any advantage to using it over either IPW or *IV(Bayes)*. The per-protocol analysis also performed well in certain scenarios, though it overestimated the treatment effect and did not maintain the type I error rate when there was a large degree of treatment effect heterogeneity and large differences in compliance between treatment arms. The per-protocol analysis also performed poorly in the re-analysis of TOPPS. As the assumptions behind the per-protocol analysis are highly likely to be violated in trials of time-varying treatments, we do not recommend its use.

The choice between IPW and *IV(Bayes)* could be made based on which set of assumptions are more plausible for a given trial. For many NI trials information on the standard treatment is available from previous studies, which could inform choice of prior for *IV(Bayes)*. However, careful consideration is required as to whether effects from previous studies will apply to the current study – if not, this could induce bias. For IPW, consideration needs to be given to potential confounders between compliance status and outcomes, and such potential confounders need to be collected during the study to be used during estimation.

In this work, we have argued that non-trial-specific intercurrent events do not pose a specific threat to the validity of NI trials, as they do not make treatment arms artificially more similar. However, such intercurrent events can still pose issues around the interpretation of results, as in any other trial. For instance, if a new treatment is only non-inferior on the basis that most patients switch to the more effective standard treatment during the trial, its use in routine care may not be warranted (even if such switching would occur in practice). Thus, in this setting it may be useful to use a hypothetical strategy for treatment switching when defining the estimand, or, alternatively, using a smaller non-inferiority margin to account for the anticipated degree of switching.

There are some limitations to this work. First, our simulation study only considered the setting with a continuous outcome and all-or-nothing compliance, and thus our results may not be generalisable to other outcome or compliance types. Second, although we generated plausible interactions in our simulation study, they may not have been sufficiently large for the *IV(interaction)* approach. Thus, our results may not apply to settings with larger interactions. Third, we only considered a frequentist evaluation of the *IV(Bayes)* method. Fourth, for IPW we did not consider methods to account for uncertainty in estimating the weights when calculating standard errors. Although we found that Stata's default sandwich estimator performed well in simulation studies, this may not be the case in other settings, and so evaluation of methods to account for such uncertainty, such as the non-parametric bootstrap, would be useful. Finally, we only considered the setting where a single binary covariate was included in IPW models. Inclusion of more variables may affect performance. For instance, IPW models dropped some observations due to perfect predictions in settings with high compliance; this issue may be exacerbated when more variables are included in the model.

The results here suggest a number of areas for future work. We have focussed primarily on defining estimands and estimators for non-inferiority trials, though it would be useful to evaluate how the estimands framework should impact on choice of non-inferiority margin. Further, as mentioned above, our simulation study focussed only on a continuous outcome with all-or-nothing compliance. It would be useful to evaluate these estimators in a wider range of settings (e.g. for binary or time-to-event outcomes, for different types of intercurrent events such as treatment switching or use of rescue medication, and for treatments which are time-varying, such as in TOPPS, rather than all-or-nothing). Further, our focus was primarily on the bias of different estimators. It may be useful to compare different approaches to calculating standard errors for each approach. IPW has been well studied in many of these settings [21], so may be a preferable option in such contexts until *IV(Bayes)* has been more fully evaluated.

## 10. Conclusions

In non-inferiority trials, trial-specific intercurrent events can make treatment arms more similar than they would be in practice, thus increasing the risk of erroneously declaring NI. To guard against this, an estimand using a hypothetical strategy for trial-specific intercurrent events should be used. IPW and *IV(Bayes)* may both be good options for estimating hypothetical effects when there is all-or-nothing compliance in two treatment arms.

**Table 1: Issue of poor adherence in non-inferiority trials from a statistical vs. estimands perspective.** ITT=intention to treat. NI=non-inferiority.

| **Statistical perspective** | **Estimands perspective** |
|---|---|
| Treatment deviations due to poor study conduct or other reasons can make treatment arms more similar than they would otherwise be. In superiority trials affected by this, ITT is "conservative" (less likely to show a statistically significant effect). However, in affected NI trials, ITT is "anti-conservative" (more likely to demonstrate non-inferiority, even when the new treatment is in fact worse). Per-protocol analyses, which exclude participants with such deviations, have been argued to be more conservative than ITT analyses in these settings (though this is not always true), and thus are often performed alongside ITT analyses to help protect against false conclusions of NI based on poor study conduct. | "Trial-specific" intercurrent events (those which occur in a trial setting but not in routine practice, for instance due to poor trial conduct, clinician decisions due to uncertainties about the evidence base, etc.) can make treatment arms more similar than they would otherwise be. Such artificial similarities between arms can lead to spurious conclusions of non-inferiority. Therefore, if trial-specific intercurrent events are likely, the estimand must account for them to avoid such spurious conclusion. A *hypothetical* strategy, which considers what outcomes would be if the trial-specific intercurrent event had *not* occurred, is a good way to do this. However, it is not always possible to distinguish between intercurrent events that would vs. would not occur outside the trial setting. Therefore, the choice of estimand in non-inferiority trials will depend both on whether trial-specific intercurrent events are likely, and whether they can be identified (see Table 2). |

**Table 2: Recommendations for applying the estimands framework to non-inferiority trials**

| | Choice of estimand |
|---|---|
| If trial-specific intercurrent events do *not* occur | A single primary estimand should be chosen based on clinical considerations, as in any other trial design. Non-inferiority should be assessed on the basis of this estimand. |
| If trial-specific intercurrent events *do* occur, and *can* be identified | A single primary estimand should be defined which handles trial-specific intercurrent events using a *hypothetical* strategy, and is otherwise defined based on clinical considerations. Non-inferiority should be assessed on the basis of this estimand. |
| If trial-specific intercurrent events *do* occur, but *cannot* be identified | Two estimands should be defined:<br>• A primary one which assumes there are no trial-specific intercurrent events and is chosen based on clinical considerations;<br>• A secondary one which uses a hypothetical strategy for any intercurrent events which may be trial-specific, as a way to protect against spurious conclusions of non-inferiority<br><br>Non-inferiority should typically be assessed on the basis of both estimands. |
| | **Choice of estimator** |
| | For trials using a hypothetical strategy to handle trial-specific intercurrent events, an estimator which targets the hypothetical estimand should be chosen (e.g. inverse probability weighting or the *IV(Bayes)* approach) with the choice depending on which assumptions are most plausible for a given trial<br><br>The assumptions behind the chosen estimator should be described, along with a discussion around the plausibility of these assumptions, and sensitivity analyses evaluating robustness of results to deviations from such assumptions if appropriate. |

**Table 3: Summary of compliance scenarios for simulation study 1**

| Scenario | Compliance type | Is association between $X/U$ and compliance the same or different between treatment arms? | Description |
|---|---|---|---|
| 1 | Compliance does not depend on any observed or unobserved baseline covariates | Same | The probability of compliance is the same for all patients |
| 2a | Compliance depends only on observed baseline covariates ($X$) | Same | Healthier patients are less likely to comply for both treatments, and there is higher compliance to the new treatment |
| 2b | | Different | Healthier patients are less likely to comply to the standard treatment, but more likely to comply to the new treatment |
| 2c | | Different | Healthier patients are less likely to comply to the standard treatment, but more likely to comply to the new treatment, and there is higher compliance to the new treatment |
| 3a | Compliance depends on both observed and unobserved baseline covariates ($X$ and $U$) | Same | Healthier patients are less likely to comply for both treatments, and there is higher compliance to the new treatment |
| 3b | | Different | Healthier patients are less likely to comply to the standard treatment, but more likely to comply to the new treatment |
| 4a | Compliance depends only on unobserved baseline covariates ($U$) | Same | Healthier patients are less likely to comply for both treatments, and there is higher compliance to the new treatment |
| 4b | | Different | Healthier patients are less likely to comply to the standard treatment, but more likely to comply to the new treatment |

**Table 4: Results from re-analysis of TOPPS trial.** ITT=intention-to-treat, PP=per-protocol, IPW=inverse probability weighting, IV=instrumental variables.

|  | Number of days with bleeding | |
|---|---|---|
|  | **Estimated difference in means (95% CI[a])** | **P-value** |
| ITT | 0.6 (0.2 to 1.0) | 0.004 |
| PP | 0.4 (-0.1 to 0.8) | 0.11 |
| IPW | 0.7 (0.2 to 1.1) | 0.005 |
| *IV(interaction)*[b] | | |
|    Relapsed disease | -1.2 (-9.7 to 7.4) | 0.79 |
|    Previous SCT | 0.0 (-15.9 to 15.9) | >0.99 |
|    Fungal infection | 3.9 (-10.2 to 17.9) | 0.59 |
|    Organ failure | 1.1 (-0.1 to 2.3) | 0.07 |
| *IV(Bayes)*[c] | | |
|    Large effect, precise | 0.5 (0.0 to 0.9) | - |
|    Small effect, precise | 0.7 (0.2 to 1.2) | - |
|    Large effect, vague | 0.5 (-0.3 to 1.3) | - |
|    Small effect, vague | 0.7 (-0.1 to 1.5) | - |

[a] CI=confidence interval for ITT, PP, IPW, IV; credible interval for Bayes

[b] Baseline characteristics used as instruments

[c] Priors were for effect of prophylaxis vs. receiving a platelet transfusion against protocol (i.e. at a higher threshold than the prophylaxis strategy calls for): large/precise~$N(2, 1)$, small/precise~$N(0, 1)$, large/vague~$N(2, 10)$, and small/vague~$N(0, 10)$.

**Figure 1: Mean estimates of treatment effect for simulation study 1 (no treatment effect heterogeneity), scenario A (true value -0.3).** ITT=intention-to-treat, PP=per-protocol, IPW=inverse probability weighting, IV=instrumental variables. *IV(interaction)* results are reported for the subset of replications where the SE was ≤10 times the SE from ITT. *IV(Bayes)* results are presented for the precise prior.

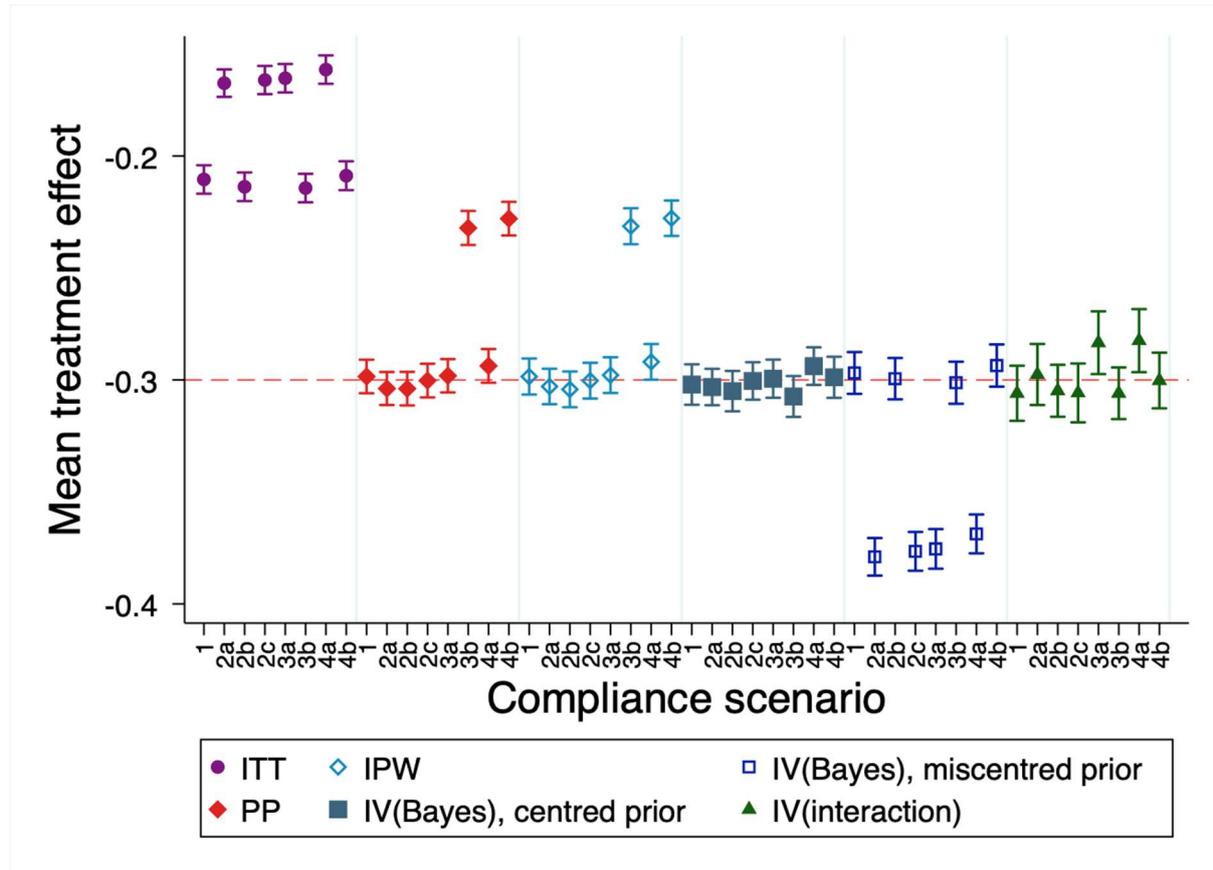

**Figure 2: Type I error rate for simulation study 1 (no treatment effect heterogeneity), scenario A (nominal value set at 2.5%).** ITT=intention-to-treat, PP=per-protocol, IPW=inverse probability weighting, IV=instrumental variables. *IV(interaction)* results are reported for the subset of replications where the SE was ≤10 times the SE from ITT. *IV(Bayes)* results are presented for the precise prior.

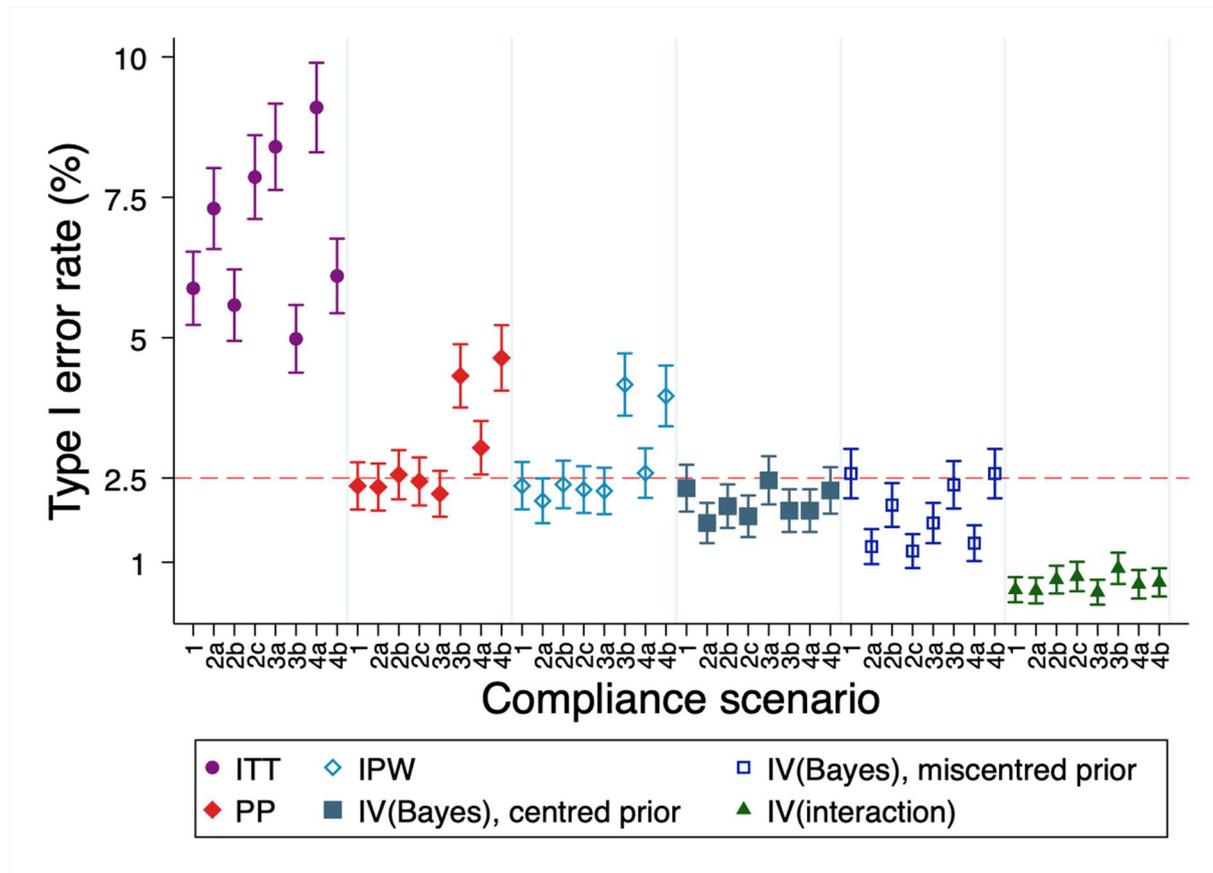

**Figure 3: Percentage increase in precision of ITT estimator vs. other estimators for simulation study 1 (no treatment effect heterogeneity), scenario A.** Defined as $100 \times \left(\left(\frac{SE_{\text{alternative}}}{SE_{\text{ITT}}}\right)^2 - 1\right)$ where $SE_{met}$ is the empirical standard error. Values >0 denote ITT is more precise than the comparator method. ITT=intention-to-treat, PP=per-protocol, IPW=inverse probability weighting, IV=instrumental variables. *IV(interaction)* results are reported for the subset of replications where the SE was ≤10 times the SE from ITT. *IV(Bayes)* results are presented for the precise prior.

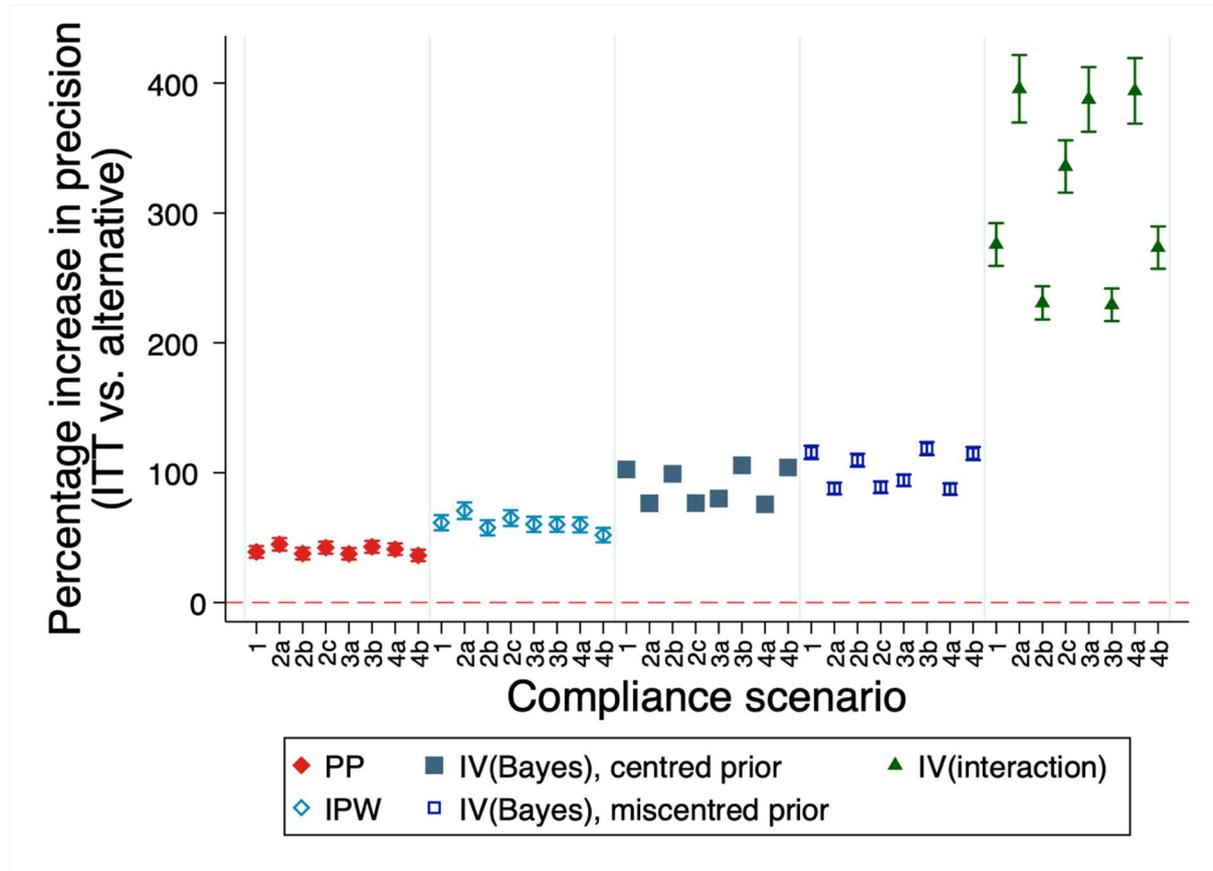

**Figure 4: Mean estimates of treatment effect for simulation study 2 (treatment effect heterogeneity) (true value -0.3).** ITT=intention-to-treat, PP=per-protocol, IPW=inverse probability weighting, IV=instrumental variables. Scenarios 1-4 relate to TEH across $X$ (an observed baseline covariate), while scenarios 5-8 relate to TEH across $U$ (an unobserved baseline covariate). Scenario 1 contains moderate compliance differences across treatment arms, and moderate TEH; scenario 2 contains large compliance differences and moderate TEH; scenario 3 contains moderate compliance differences and large TEH; and scenario 4 contains large compliance differences and large TEH. A similar pattern occurs for scenarios 5-8. *IV(interaction)* results are reported for the subset of replications where the SE was ≤10 times the SE from ITT.

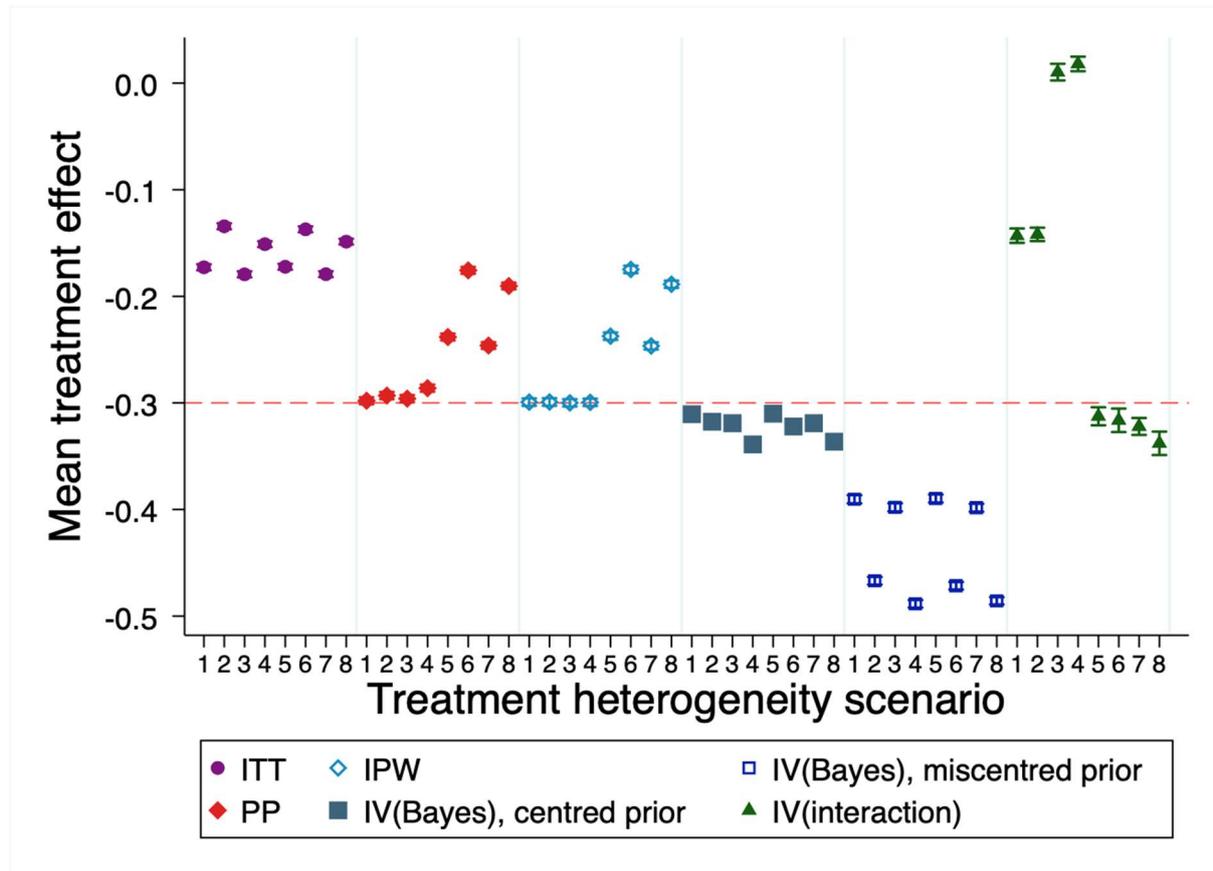

**Funding**
BCK and IRW are funded by the UK MRC, grants MC_UU_00004/07 and MC_UU_00004/09. KEM is funded by a UK MRC Skills Development Fellowship MR/P014372/1. CL is funded by a UK MRC Skills Development Fellowship MR/T032448/1.

**Contributions**
KEM, IRW, CL, and BCK designed the simulation study. KEM implemented the simulation study. BCK re-analysed the TOPPS data. KEM and BCK wrote the first draft of the manuscript. IRW, CL, and SS revised the manuscript. All authors read and approved the final manuscript.

**Declaration of Conflicting Interests**
The Authors declare that there is no conflict of interest.

**Data availability**
Data sharing requests for the TOPPS data must be directed towards the trial sponsor.